\documentclass{elsart3}

\usepackage{graphicx}

\usepackage{amssymb}
\usepackage{amsmath}
\usepackage{subfigure}

\input symbols
\def\bb{\ensuremath{\Bs {\kern -0.16em \Bsb}}\xspace}
\def\b    {\ensuremath{B^0_s}\xspace}
\begin{document}

\begin{frontmatter}



\title{Neutrino reconstruction with topological information}

\author[eth,psi]{S. Dambach},
\author[eth]{U. Langenegger}, 
\author[eth]{A. Starodumov\corauthref{cor1}}\ead{Andrey.Starodumov@cern.ch},
\address[eth]{Institute for Particle Physics, ETH Zurich, 8093 Zurich, Switzerland}
\address[psi]{Paul Scherrer Institute, 5232 Villigen PSI, Switzerland}
\corauth[cor1]{Corresponding author}

\begin{abstract}

In general a decay with a missing (not detected) particle can not be
fully reconstructed apart from a few exceptions. For example, if the
momentum of the decaying particle is known or if the missing energy in
an event is measured precisely, then the missing particle 4-momentum
can be determined.  Here a new method is proposed that utilizes
additional information about the topology of a decay: The direction
from the primary to the secondary vertex combined with momentum
conservation 
allows the
determination of the missing particle momentum up to a twofold
ambiguity. The semileptonic decay of the \Bs\ mesons is considered as an
example to illustrate this method and to compare its performance against
conventional approaches.

\end{abstract}

\end{frontmatter}

\section{Introduction}
\label{introduction}


Generally it is considered that the decay of a particle, where one of
the final state products is not detected, can not be fully reconstructed.
There are a few well known exceptions. If the momentum of the decaying
particle is known and all but one of the decay products are detected
and reconstructed then, obviously, it is possible to determine the
missing particle's 4-momentum. Another possibility to measure the
momentum of missing particle is based on the hermeticity of the
detector where the missing energy is measured sufficiently well.
An approximate method is used in studies of $\bb$ oscillations with
semileptonic decays where the neutrino is a missing particle. The
reconstruction of the $\b$ meson momentum is a key ingredient in these studies. 
For this purpose, a $k$-factor is introduced which is the ratio
between the momentum of the measured decay products and the original
$\b$ momentum. The $k$-factor is obtained from Monte Carlo (MC)
studies.  The use of the $k$-factor induces a large error
on the proper time of the $\b$.
  
In this paper we point out that by including additional topological
information the neutrino 4-momentum can be reconstructed up to a
twofold ambiguity. The study of \bb - oscillations in the semileptonic decay of the \b meson at hadron
colliders is used to illustrate the method of the neutrino 4-momentum
reconstruction.

\section{Kinematics of the Semileptonic \Bs\ Decay}
\label{kinematics}
Consider the decay $\Bs \rightarrow \Dsm \ellp \nu$ where the \Dsm
meson decays into a final state with charged hadrons and all hadrons
are reconstructed. The kinematics of such a decay can be described by
four equations in three dimensions:

\begin{equation}
\sqrt{m_B^2+\vec{P}_B^2} = \sqrt{m_{(D_s \ell)}^2+\vec{P}_{(D_s \ell)}^2} + |\vec{P}_{\nu}|
\end{equation}

\begin{equation}
\vec{P}_B = \vec{P}_{(D_s \ell)} + \vec{P}_{\nu}
\end{equation}

In this system of equations there are six unknown variables: $P_B^i$,
$P_{\nu}^i$, where $i$ stands for $x$, $y$ and $z$. Since there are only
four equations, the system is undetermined. 

The flight direction of the \Bs\ meson, $\vec{V}_B$, obtained from the primary and
secondary vertex, provides the necessary
information to solve for the neutrino momentum. Without loss of
generality one can work in a two dimensional coordinate system: one 
axis ($\parallel$) is defined by the \Bs\ flight direction $\vec{V}_B$,
the other one ($\perp$) is defined in the plane spanned
by the two vectors $\vec{P}_{\Ds\ell}$ and $\vec{V}_B$ and is
orthogonal to the \Bs\ flight direction. $\Theta$ is the angle between the
flight directions of the \Bs meson and the $D_s \ell$ system. Then one gets:

\begin{eqnarray}
\sqrt{m_B^2+\vec{P}_B^2} &=& \sqrt{m_{(D_s \ell)}^2+ \vec{P}_{(D_s \ell)}^2} + |\vec{P}_{\nu}| \\
|\vec{P}_B| &=& P_{(D_s \ell)}^{\parallel} + P_{\nu}^{\parallel} \\
\vec{P}_{\nu}^2 &=& (P_{\nu}^{\parallel})^2 + (P_{\nu}^{\perp})^2\\
\vec{P}_{(D_s \ell)}^2 &=& (P_{(D_s \ell)}^{\parallel})^2 + (P_{(D_s \ell)}^{\perp})^2\\
P_{(D_s \ell)}^{\parallel} &=& |\vec{P}_{(D_s \ell)}| \times \cos{\Theta} \\
\cos{\Theta} &=& \frac{\vec{P}_{(D_s \ell)} \cdot \vec{V}_B}{|\vec{P}_{(D_s \ell)}|  |\vec{V}_B|}
\end{eqnarray}

The momentum components of the $\Ds\ell$ system and of the
neutrino orthogonal to the \Bs\ flight direction are of equal magnitude and opposite sign:

\begin{eqnarray}
P_{(D_s \ell)}^{\perp} &=& -P_{\nu}^{\perp}
\end{eqnarray}

Therefore the neutrino momentum component along the \Bs\ flight
direction can be obtained with a straightforward calculation:

\begin{eqnarray}
P_{\nu}^{\parallel} &=& -a \pm \sqrt{r}
  \label{eq:nu}
\end{eqnarray}

where

\begin{eqnarray}
a &=&  \frac{(m_B^2 - m_{(D_s \ell)}^2 - 2 \cdot (P_{(D_s \ell)}^{\perp})^2) \cdot
  P_{(D_s  \ell)}^{\parallel}} {2 \cdot ((P_{(D_s  \ell)}^{\parallel})^2 - E_{(D_s \ell)}^2)}\\
r &=& \frac{(m_B^2 - m_{(D_s \ell)}^2 - 2 \cdot (P_{(D_s \ell)}^{\perp})^2)^2 \cdot E_{(D_s \ell)}^2}{4 \cdot ((P_{(D_s  \ell)}^{\parallel})^2 - E_{(D_s \ell)}^2)^2}
+ \frac{E_{(D_s \ell)}^2 \cdot (P_{(D_s \ell)}^{\perp})^2}{(P_{(D_s  \ell)}^{\parallel})^2 - E_{(D_s \ell)}^2}
\end{eqnarray}



\hspace{1cm}

Eq.~\ref{eq:nu} leads to two solutions, only one of them is correct. In an experimental environment there might be no solution of this equation if the radicand $r$ is negative due to finite vertex and momentum resolutions. In the following we illustrate that despite these facts the method still provides competitive results with respect to conventional approaches.   




\section{MC simulations }
\label{toyMC}

To verify that the proposed method works not only in ideal conditions
without experimental errors, a MC simulation has been developed to study
\bb\ - mixing in the semileptonic decay mode.
The PYTHIA V6.227 \cite{Sjostrand:2006za} package has been used as an
event generator with a center-of-mass energy of 14~TeV.
We have generated 45000 events each containing two b~quarks. One of them hadronizes
to a \Bs\ meson and decays according to
\begin{equation}
\Bs\ \rightarrow \Ds \mu^+ \nu_{\mu}, \qquad \Ds\rightarrow \phi \pi^-,\qquad \phi \rightarrow K^+ K^-; 
  \label{eq:decay}
\end{equation}
the second $b$ quark hadronizes according to PYTHIA into a
$b$-hadron and decays unconstrained. The decay in (\ref{eq:decay})
is called `signal' decay to distinguish it from the other $b$-hadron decay.

All hadrons from the signal decay are required to have $\pt \ge 1\gevc$,
 the muon is required to have
$p_T \ge 3\gevc$. The track parameters and vertex positions (primary
and secondary) have been smeared according to Gaussian distributions with the following parameters.
The momentum uncertainty has been simulated by smearing the pseudorapidity with $\sigma_{\eta} =
5.8\times 10^{-4}$, the angle $\phi$ with $\sigma_{\phi} = 0.58\mrad$
and the inverse transverse momentum  with  $\sigma_{(1/p_T)} = 0.013
 (GeV/c)^{-1}$. 
The primary vertex has been smeared with $\sigma_{x,y} = 20\mum$ in
both, x- and y-direction, the secondary vertex has been smeared with
$\sigma_{||} = 70\mum$ in flight direction of the \Bs\ and $\sigma_{\perp} =
10\mum$ in the perpendicular direction.



\section{Proper time reconstruction}
\label{ptime}

The most important ingredient in the measurement of the \Bs oscillation frequency is the proper time
calculated in the transverse plane\footnote{Here the
  transverse plane is perpendicular to the beam direction, as is usual
in collider experiments.} as follows:
\begin{equation}
  \label{eq:pt}
  c\tau = \frac{L_{xy} \, m(\Bs)}{p_T(\Bs)}, 
\end{equation}
where $L_{xy}$ is the flight path length, $m(\Bs)$ and $p_T(\Bs)$ are
the mass and transverse momentum of the \Bs\ meson respectively.

In the semileptonic $\b$ meson decay, the neutrino is not detected and
the \pt\ of the \Bs\ meson cannot be measured directly. Therefore a
correction factor, derived from MC simulations, is introduced to scale
the measured transverse momentum of the $\Ds\ell$
system. Eq.~\ref{eq:pt} is modified as follows
\begin{equation}
  \label{eq:ptk}
  c\tau = \frac{L_{xy}\,  m(\Bs)}{p_T(D_s \ell)} \times k,
\end{equation}
with the $k$-factor estimated from MC simulations and calculated as
\begin{equation}
k = \frac{p_T(D_s\ell)}{p_T(\Bs)}. 
\end{equation}

\begin{figure}[hbt]
  \begin{center}
    \includegraphics[scale=.39]{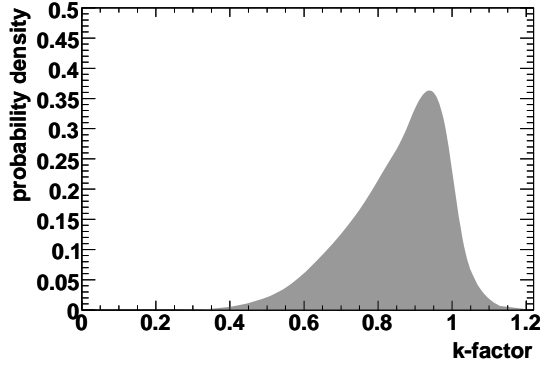}
    \caption{Probability density for the $k$-factor.}
  \label{fig:kFactor}
  \end{center}
\end{figure}

\begin{figure}[hbt]
  \begin{center}
    \includegraphics[scale=.39]{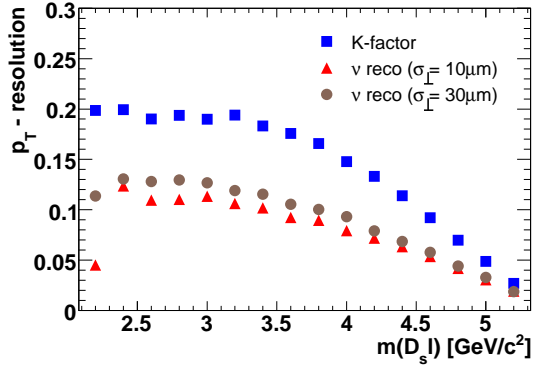}
    \caption{\Bs\ momentum resolution calculated with $k$-factor and neutrino reconstruction methods.}
  \label{fig:ptres1}
  \end{center}
\end{figure}

This $k$-factor (see Fig.~\ref{fig:kFactor}) introduces a significant error on the \Bs\ momentum
and, hence, on the proper time as well. 
To reduce the error by an average $k$-factor, it is calculated in bins
of $m_{(D_s \ell)}$. We illustrate in Fig.~\ref{fig:ptres1} the \Bs\
meson transverse momentum resolution as a function of the invariant mass
of the $\Ds\ell$ system for the $k$-factor and the neutrino reconstruction
method. It is evident that the resolution obtained
with the neutrino reconstruction method is substantially better than
with the $k$-factor method. Note that the neutrino reconstruction method is
sensitive to $\sigma_{\perp}$, therefore two different performances for
$\sigma_{\perp} = 10 \mum$ and $\sigma_{\perp} = 30 \mum$ are shown. 

\begin{figure}[hbt]
  \begin{center}
    \unitlength1.0cm 
    \begin{picture}(25., 6.)
      \put(  0.0,  0.) {\includegraphics[scale=.39]{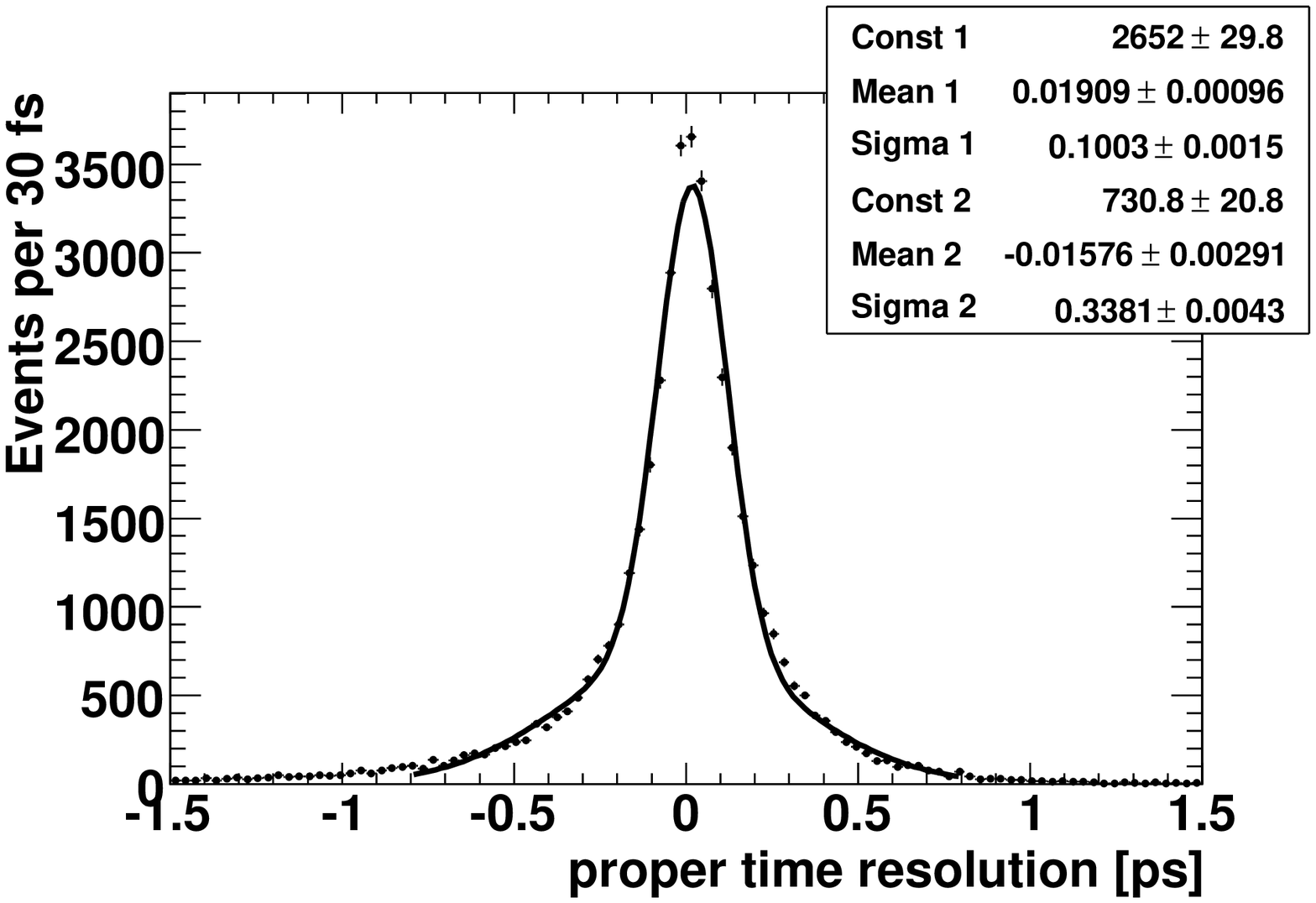}}
      \put(  8.5,  0.) {\includegraphics[scale=.39]{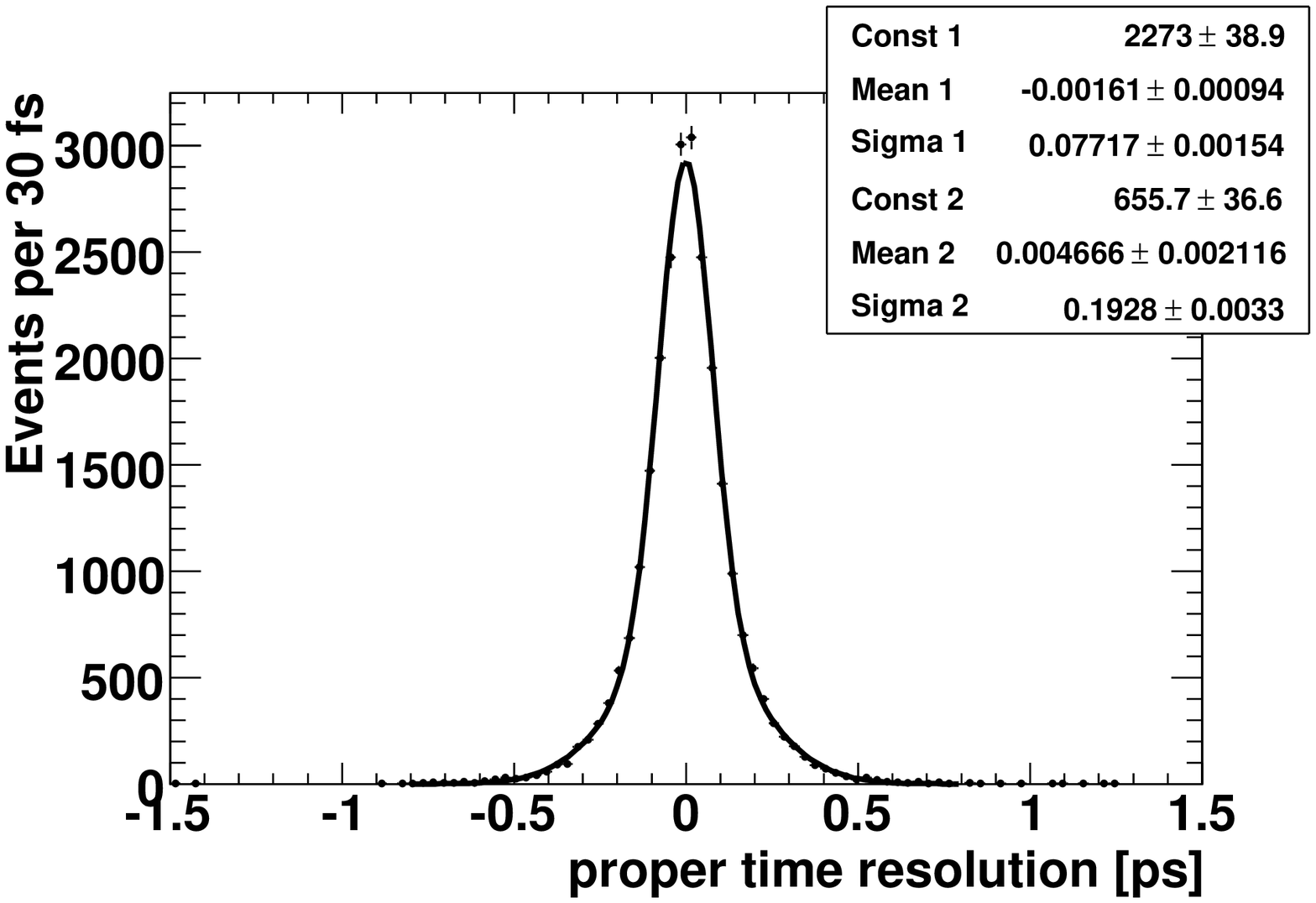}}
      \put(  1.4,  4.2){(a)}
      \put(  9.9,  4.2){(b)}
    \end{picture}
    \caption{Proper time resolution.  (a)  $k$-factor
      method, (b) neutrino reconstruction method. }
  \label{fig:taures}
  \end{center}
\end{figure}

Fig.~\ref{fig:taures} illustrates the proper time resolution obtained
with the $k$-factor method (a) and  the neutrino
reconstruction method (b). For the latter method the closest to the
true value solutions are filled in the histogram.
The distributions are fitted with two Gaussian, 
the average width $\sigma$ is determined according to

\begin{equation}
 \sigma^2 = \frac{N_n^2\sigma_n^2 + N_w^2\sigma_w^2}{N_n^2 + N_w^2},
\end{equation}

where $\sigma_n$ ($\sigma_w$)
and $N_n$ ($N_w$) are the width
and normalization of the narrow (wide) Gaussian, respectively.
For the $k$-factor method  $\sigma_{n} = 100\fs$ ($\sigma_{w} = 338\fs$)
and  $N_n=2700$ ($N_w=730$) that gives an average  $\sigma = 132\fs$.
For the neutrino reconstruction method  $\sigma_{n} = 77\fs$ ($\sigma_{w} = 193\fs$)
and  $N_n=2300$ ($N_w=660$) that gives an average  $\sigma = 91\fs$.
The neutrino reconstruction method provides a proper time resolution which 
is better than in the reconstruction done with 
the $k$-factor method.

\section{Amplitude Fit}
\label{example}

A standard way to search for \Bs\ oscillations is the amplitude method
described in Ref.~\cite{Moser:1996xf}. Briefly it consists of the following
steps. Candidates for the
decay~(\ref{eq:decay}) are split into two samples.
Events with the same flavor
at production and decay time comprise the unmixed sample, events with
the opposite flavor constitute the mixed sample. The two cases
are distinguished by tagging the flavor of the \Bs\ meson at the
production and the decay. Tagging at the production  can be achieved
by identifying the flavor of the second $b$-hadron in the event;
tagging at the decay can be accomplished by identifying the
charge of the \Bs\ meson decay products.

In each event, the proper decay time of the \Bs\ mesons is
reconstructed and the two event samples are used to define the
time-dependent asymmetry
\begin{equation}
a(t) = \frac{P_{unmix}-P_{mix}}{P_{unmix}+P_{mix}} \propto A \times D \times \cos(\Delta m_s t),
  \label{eq:as}
\end{equation}
where $D$ is a global dilution factor accounting for background,
miss-tagging and proper-time resolution and $A$ is the amplitude. 
$P_{unmix}$ and $P_{mix}$ are the time-dependent probability
distribution functions for mixed and unmixed \Bs\ meson decays,
respectively. 


In the fit to the asymmetry distribution, the oscillation frequency
$\Delta m_s$ is fixed, leaving the amplitude $A$ as a free
parameter. A scan over $\Delta m_s$ is performed starting from
zero. If the fixed value of $\Delta m_s$ is consistent with the true
one, the fitted amplitude will be equal to unity, else it is
consistent with zero. The error on the amplitude value $A$ is calculated
according to the formula

\begin{equation}
\sigma_A = \frac{1}{1-2W} \times \sqrt{\frac{2}{S+B}} \times \frac{S+B}{S} \times e^{\textstyle \frac{\Delta m_s^2 \sigma_t^2}{2}},   
\end{equation}
where $W$ is the mistagging probability, $S$ the number of signal, $B$
the number of background events, and $\sigma_t$ the proper time
resolution. The amplitude method is sensitive to the tested value of
$\Delta m_s$ if $\sigma_A$ is small compared to unity. 
The value of $\Delta m_s$ at which $1 = 1.645 \times \sigma_A $ is quoted as
sensitivity limit.
 
In the following we assume a mistagging
probability of $40\%$, a number of 45000 signal events for semileptonic
\Bs\ decays and a signal to background ratio of 1:1. The
simulated oscillation frequency is taken as 
$\Delta m_s=17.25~\mathrm{ps^{-1}}$~\cite{Abulencia:2006mq,Abazov:2006dm}.

\begin{figure}[hbt]
  \begin{center}
    \unitlength1.0cm 
    \begin{picture}(25., 6.)
      \put(  0.0,  0.) {\includegraphics[scale=.39]{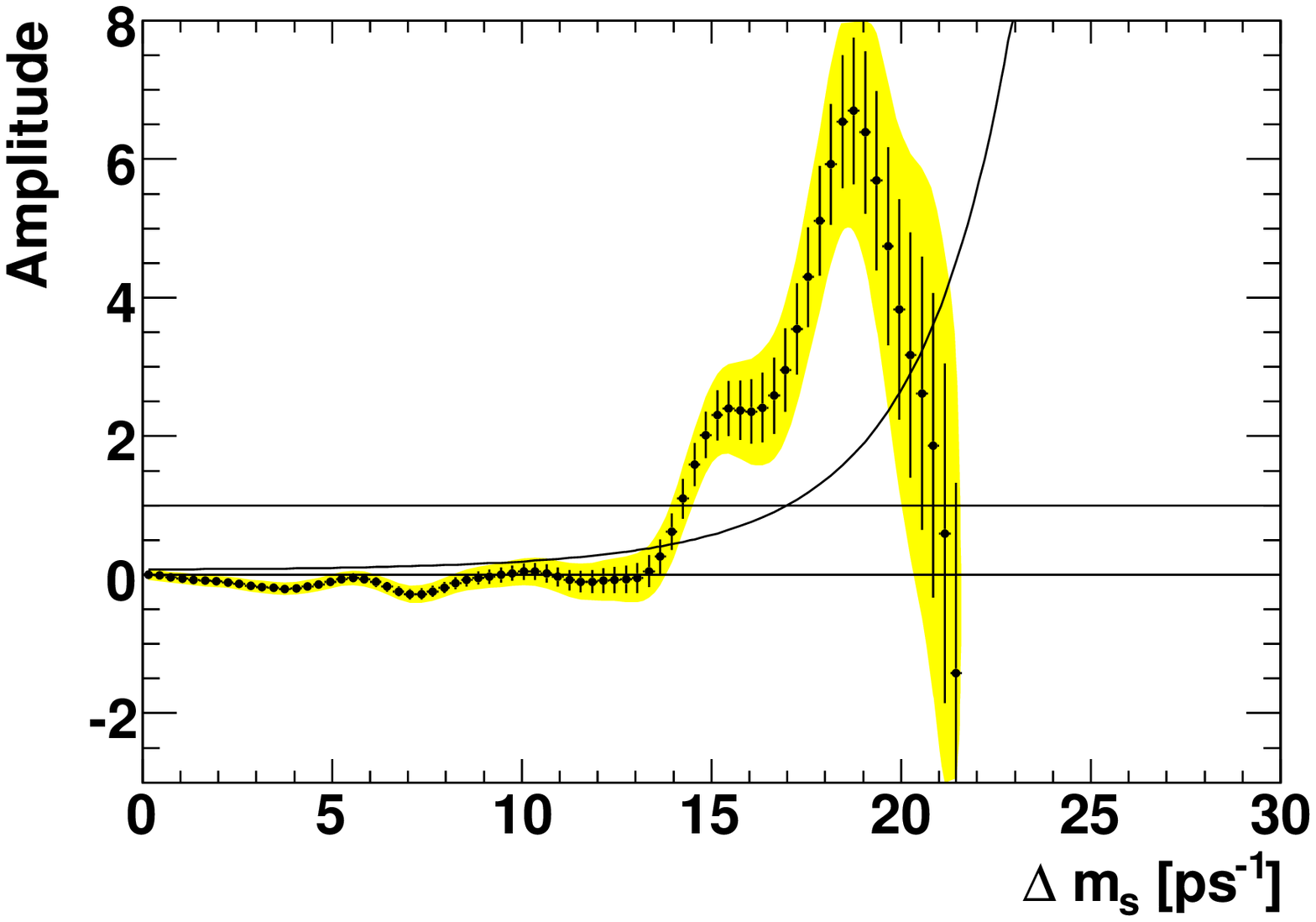}}
      \put(  8.5,  0.) {\includegraphics[scale=.39]{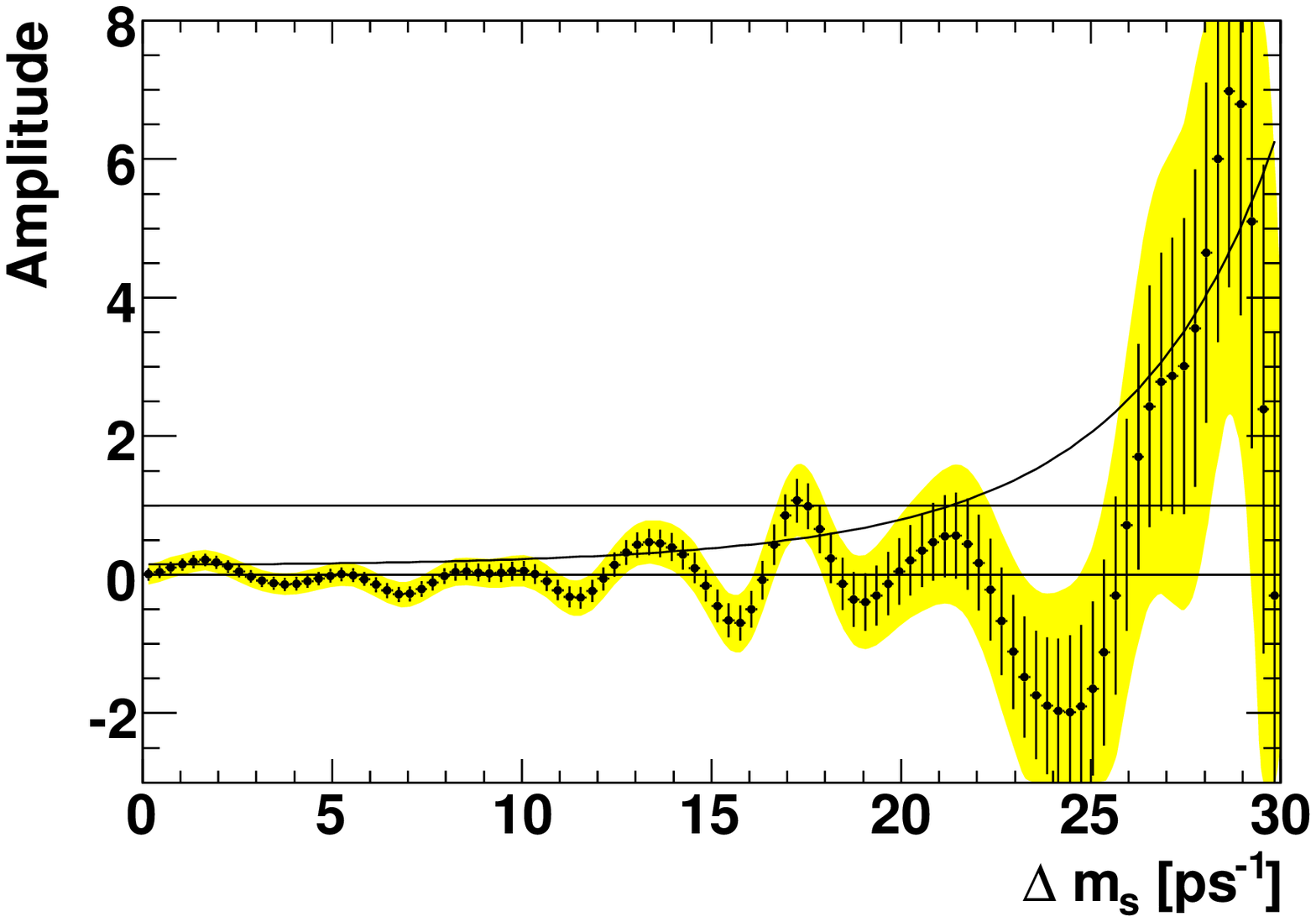}}
      \put(  1.4,  4.2){(a)}
      \put(  9.9,  4.2){(b)}
    \end{picture}
    \caption{Amplitude fits for \bb\ oscillations. (a)  $k$-factor
      method, (b) neutrino reconstruction method. }
  \label{fig:ampscan}
  \end{center}
\end{figure}

The result of the amplitude fit for the $k$-factor
method with the sensitivity curve is
presented in Fig.~\ref{fig:ampscan}a. The sensitivity of the
method for the processed number of signal events is about 17~$\mathrm{ps^{-1}}$.

On Fig.~\ref{fig:ampscan}b one can see the result of the amplitude fit
done on the events reconstructed with the neutrino reconstruction
method.  The sensitivity to $\Delta m_s$  is about 21~$\mathrm{ps^{-1}}$. 
In this case there is a clear peak in the amplitude fit at the input value
of $\Delta m_s=17.25~\mathrm{ps^{-1}}$.

\section{Discussion}
\label{discussion}
On the one hand the use of the neutrino reconstruction method improves the momentum resolution 
of the \Bs\ meson. On the other hand it degrades the signal to background ratio and
reduces the signal sample.
First, since solving the system of equations, we end up with a
quadratic equation with two solutions. One solution is the `true',
the other is the `wrong' one. In such a way the number of background
events is increased by the number of wrong solutions from the signal as well as
from the background sample.
Second, again due to the quadratic equation and the realistic resolutions,
a certain fraction of events has a negative radicand $r$ and, hence,
there is no solution for the \b meson momentum. These events are
excluded from the analysis (both from signal and background samples).

With the resolutions used in this analysis the number of signal events is
decreased by a factor of 2 roughly (the same is true also for
background) and the number of background events is increased by a factor of 3 (due to the
`wrong' signal and background events). Hence, in the neutrino
reconstruction method the signal to background ratio is not 1:1 (like
in the $k$-factor method) but 1:3. Despite these facts the sensitivity of the
neutrino reconstruction method is still at higher values of $\Delta m_s$ compared to
the $k$-factor method.

\begin{figure}[hbt]
  \begin{center}
    \unitlength1.0cm 
    \begin{picture}(25., 6.)
      \put(  0.0,  0.) {\includegraphics[scale=.39]{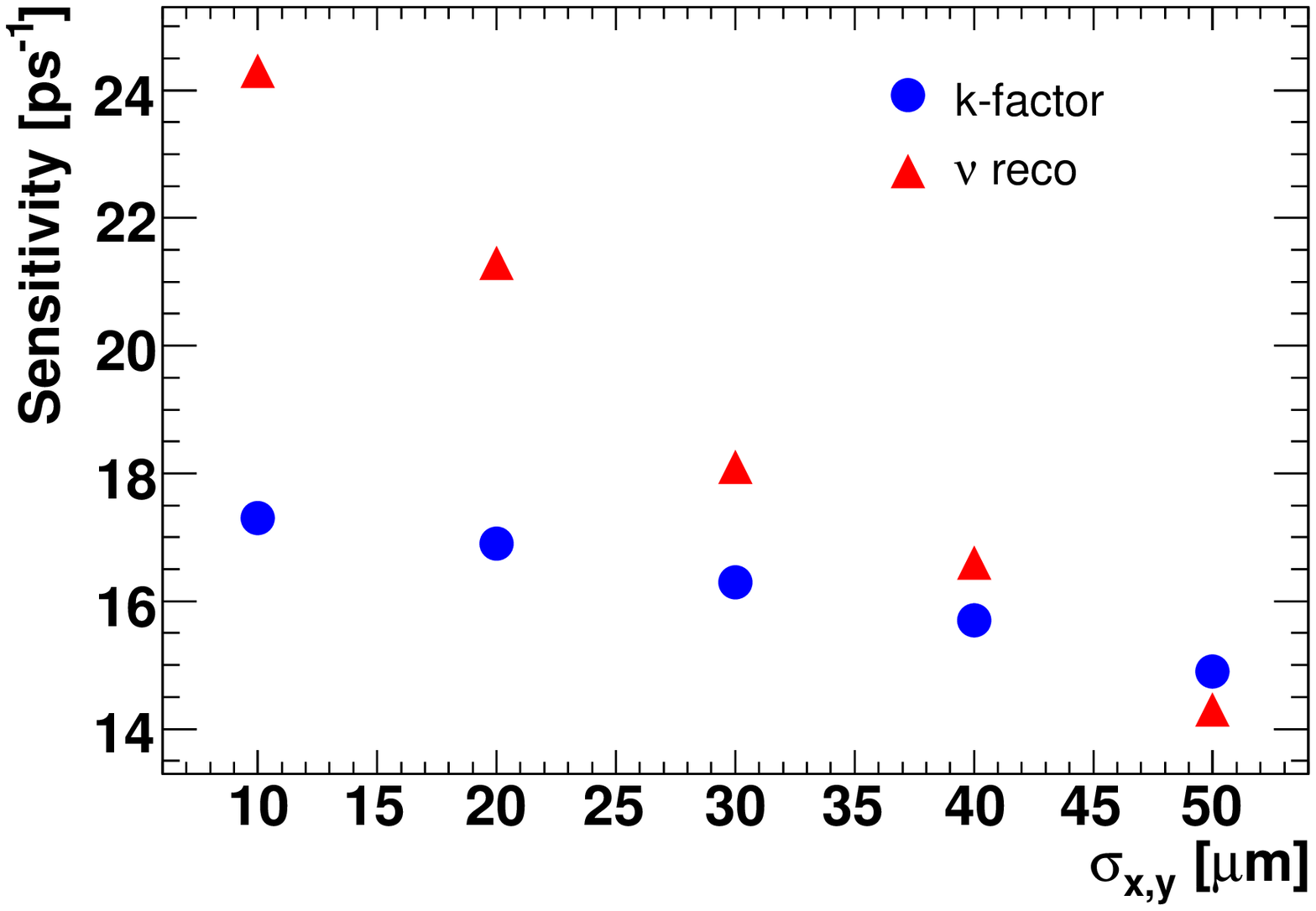}}
      \put(  8.5,  0.) {\includegraphics[scale=.39]{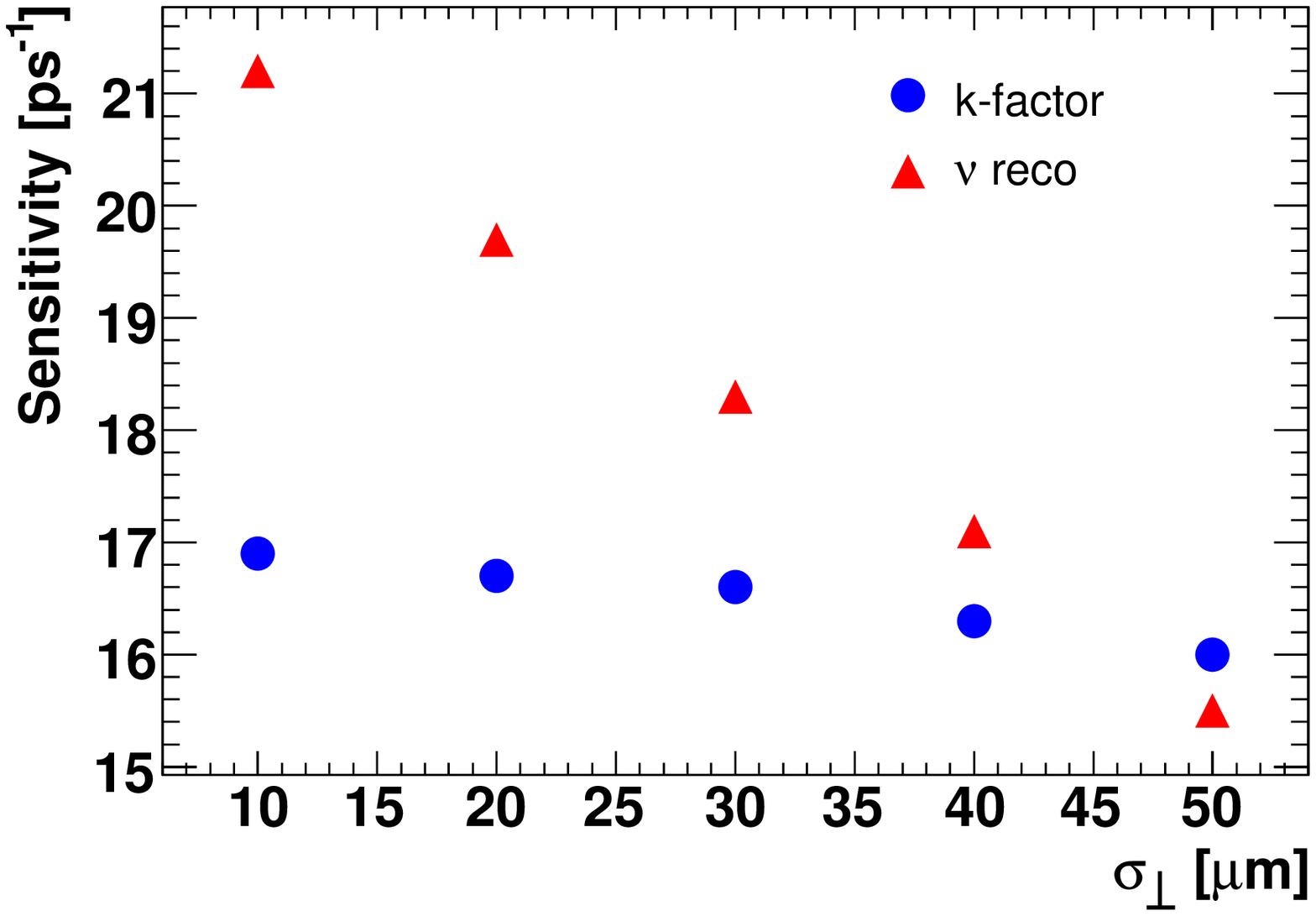}}
      \put(  6.2,  4.0){(a)}
      \put(  14.7,  4.0){(b)}
    \end{picture}
    \caption{Sensitivity as a function of (a) the primary vertex resolution $\sigma_{x,y}$  and
      (b) the secondary vertex resolution $\sigma_{\perp}$ for neutrino
      reconstruction and $k$-factor methods.}
    \label{fig:compknures}
  \end{center}
\end{figure}

Also the quality of both methods with respect to the primary
(Fig.~\ref{fig:compknures}a) and secondary (Fig.~\ref{fig:compknures}b)
vertex resolution is investigated, when all other resolutions are kept constant.
From these 
figures one can conclude that the neutrino reconstruction method is more powerful than 
the $k$-factor method if $\sigma_{\perp} \lesssim 45\mu$m  and $\sigma_{x,y} \lesssim 45\mu$m. 
For a typical hadron collider detector resolutions better than quoted above
are achievable~\cite{Starodumov:1997,atlas:1999fq,Abe:1998cj,Abazov:2006cb}.

\section{Conclusion}
\label{conclusion}

In this paper we have shown that the full neutrino reconstruction is
possible in decays like the semileptonic $\b$ decay mode where all but one final state particle are
measured. For this one has to use additional topological information, in our example the
direction of the $\b$ meson momentum. The search for the $\bb$ oscillations has
been used as an illustration of the method. For the
verification of our procedure we have used typical resolutions of 
hadron collider detectors. The sensitivity obtained with the 
neutrino reconstruction method is much higher than with 
the convenient method used for this decay mode.


Finally, we would like to stress that the
proposed method can be used not only in semileptonic \B\ decays but 
in some other cases where the known topology of a decay can compensate for the incompleteness 
of kinematical information.

\section{Acknowledgment}
\label{acknowledgment}

We would like to thank our colleagues from PSI for fruitful discussions. This research  was supported by the Swiss  National Science Foundation (SNF).

\end{document}